\documentclass[square,comma,sort&compress]{elsart}
\usepackage{amsfonts}
\usepackage{amsmath}
\usepackage{amssymb}
\usepackage{natbib}
\usepackage{epsfig}
\usepackage{rotating}
\usepackage{graphicx}

\newcommand{\QUESTION}[1]{}
\newcommand{\ANSWER}[1]{}

\begin{document}

\runauthor{Ellero}
\begin{frontmatter}
\title{An H-theorem for incompressible fluids}
\author[UTS]{M. Tessarotto}
\author[SYD]{and M. Ellero}

\address[UTS]{Department of Mathematics
and Informatics, University of Trieste, Trieste, Italy and
Consortium for Magnetofluid Dynamics, Trieste, Italy}

\address[SYD]{School of Aerospace, Mechanical and Mechatronic Engineering, The
University of Sydney, NSW 2006, Australia}

%
\thanks[marco]{Corresponding author: email: Marco.Ellero@cmfd.univ.trieste.it}

\begin{abstract}
A basic  aspect of the kinetic descriptions of incompressible fluids based on an inverse
kinetic approaches is the possibility of satisfying an H-theorem. This property is in fact
related to the identification of the kinetic distribution function with a probability density in an
appropriate phase space. Goal of this investigation is to analyze the conditions of validity of
an H-theorem for the inverse kinetic theory recently proposed by  Ellero and Tessarotto
[2004, 2005]. It is found that the time-dependent contribution to the kinetic pressure,
characteristic of such a kinetic model, can always be uniquely defined in such a way to
warrant the constancy of the entropy.
\newline


\end{abstract}

\end{frontmatter}

\section{Introduction}

The possibility of defining an inverse kinetic theory for the 3D
incompressible Navier-Stokes equations (INSE; see Appendix A), recently
pointed out \cite{Ellero2000,Tessarotto2004,Ellero2005} raises the
interesting question whether the relevant kinetic distribution function
satisfies an H-theorem, namely the related (Shannon) kinetic entropy can be
specified in such a way to result monotonically non-decreasing in time for
arbitrary fluid fields defined in a internal domain, to be identified with a
bounded three-dimensional domain $\Omega \subseteq
\mathbb{R}
^{3}$ (fluid domain). The validity of such a theorem is in fact a sufficient
condition of strict positivity of the distribution function. Therefore, this
result is important in order to establish also the consequent interpretation
of the kinetic distribution function, to be suitably normalized, in terms of
a probability density on an appropriate phase space $\Gamma $.

Purpose of this Note is to evaluate the kinetic entropy and the related
entropy production rate for strictly positive, suitably smooth, but
otherwise arbitrary, distribution functions $f(\mathbf{x,}t)$ which
correspond to an arbitrary strong solution of the initial-boundary value
problem of INSE (see Appendix A and Refs.\cite{Ellero2005,Ellero2006}),
defined for an internal domain of $%
\mathbb{R}
^{3},$\ and for kinetic distributions functions which are not necessarily
Maxwellian.\textit{\ \emph{We intend to prove that, under a suitable
assumptions, which involve the specification of the only non-observable free
parameter of the theory, a strictly positive time-dependent additive
contribution to the kinetic pressure, the kinetic entropy results
identically conserved, thus yielding an H-theorem for the kinetic
distribution function.}\ }The conclusion is obtained invoking mild
assumptions both on the initial kinetic distribution function and on the
fluid fields $\left\{ \rho (\mathbf{r},t)=\rho _{o}>0,\mathbf{V}(\mathbf{r}%
,t),p(\mathbf{r},t)\right\} $. In the sequel we intend to show that the
proof of this statement, and hence of the positivity of the kinetic
distribution function, relies essentially only on the following assumptions:

a) the strict positivity of the initial distribution function $f(\mathbf{x,}%
t_{o})$ in the whole phase space $\Gamma ;$

b) the requirement that $f(\mathbf{x,}t_{o})$ belongs to the functional class%
\begin{equation}
f(\mathbf{x,}t_{o})\in C^{(1,1)}(\Omega \times I),  \label{smoothness of f}
\end{equation}%
and results suitably summable in $\Gamma $ and smooth in $\Gamma \times I;$

c) no-slip, Dirichlet boundary conditions are imposed on the fluid fields
(see Appendix A);

d) the total mass of the fluid is conserved, i.e.,%
\begin{equation}
\int\limits_{\Omega }d^{3}\mathbf{r}\rho (\mathbf{r},t)=M=const.
\label{INSE-6c}
\end{equation}

e) the fluid is subject to a volume force density $\mathbf{f(r,}t)$ which is
assumed suitably smooth, precisely at least
\begin{eqnarray}
\mathbf{f}(\mathbf{r,}t) &\in &C^{(1,0)}(\Omega \times I),  \label{INSE-7} \\
\mathbf{\mathbf{f}}(\mathbf{\mathbf{r,v}},t\mathbb{)} &\in &C^{(0)}(%
\overline{\Omega }\times I),  \label{INSE-8}
\end{eqnarray}%
where $C^{(i,j)}(\Omega \times I)\equiv C^{(i,)}(\Omega )\times C^{(j)}I),$
with $i,j\in
\mathbb{N}
;$

f) the existence of a strong solution of INSE exists in $\overline{\Omega }%
\times I$ which belongs to the functional class:\bigskip
\begin{equation}
\left\{
\begin{array}{l}
\mathbf{V}(\mathbf{\mathbf{r,}}t),p(\mathbf{r,}t)\in C^{(0)}(\overline{%
\Omega }\times I), \\
\mathbf{V}(\mathbf{\mathbf{r,}}t),p(\mathbf{r,}t)\in C^{(2,1)}(\Omega \times
I),%
\end{array}%
\right.  \label{NFS}
\end{equation}%
with $I$ denoting the time axis, generally to be identified with a finite
subset of $%
\mathbb{R}
;$

g) the determination of the time-dependent term of the kinetic pressure ($%
p_{1}$). It if found that $p_{1}$ can always be defined in the same time
interval $I$, up to an arbitrary positive constant, in such a way that the
entropy production rate vanishes identically.

The basic result can be summarized as follows.

\begin{thm}
\textbf{- H-theorem} \textbf{(entropy conservation for the inverse kinetic
theory of \ INSE)}
\end{thm}

\emph{Let us denote by} $f(\mathbf{x}(t)\mathbf{,}t)$ \emph{the kinetic
distribution function solution, assumed to exist and result suitably
regular, of the inverse kinetic equation represented in the form}
\begin{equation}
f(\mathbf{x}(t)\mathbf{,}t)=T_{t,t_{o}}f(\mathbf{x}_{o}\mathbf{,}t_{o}),
\label{eq.n-1}
\end{equation}%
$\mathbf{x}=(\mathbf{r,v})$ \emph{being a state vector belonging to the
phase space }$\Gamma =\Omega \times U$ \emph{and\ }$U=%
\mathbb{R}
^{3}$\emph{\ the velocity space. }$f(\mathbf{x}_{o}\mathbf{,}t_{o}),$ \emph{%
the initial kinetic distribution function, is assumed to be suitably smooth
in the sense of Eq.(\ref{smoothness of f}), strictly positive and summable
in }$\Gamma $ \emph{for} \emph{appropriate weight functions. Moreover, }$%
T_{t,t_{o}}$\emph{\ is a suitably-defined diffeomorphism, denoted as
Navier-Stokes evolution operator, such that }$\forall t\in I,$ \emph{%
including the initial time }$t_{o}\in I,$ \emph{the fluid fields at any time
}$t\in I,$ $\left\{ \rho _{o},\mathbf{V}(\mathbf{r},t),p(\mathbf{r}%
,t)\right\} $ \emph{are uniquely defined by the following velocity moments
of the kinetic distribution function:}%
\begin{equation}
\rho _{o}=\int d^{3}vf(\mathbf{x,}t)=\rho _{o}=\int d^{3}vT_{t,t_{o}}f(%
\mathbf{x}_{o}\mathbf{,}t_{o}),  \label{1m}
\end{equation}%
\begin{equation}
\mathbf{V(r,}t)=\frac{1}{\rho _{o}}\int d^{3}v\mathbf{v}f(\mathbf{x,}t)=%
\frac{1}{\rho _{o}}\int d^{3}v\mathbf{v}T_{t,t_{o}}f(\mathbf{x}_{o}\mathbf{,}%
t_{o}),  \label{2m}
\end{equation}%
\begin{equation}
p\mathbf{(r,}t)=p_{1}\mathbf{(r,}t)-P_{o},  \label{3m}
\end{equation}%
\emph{where}%
\begin{equation}
p_{1}\mathbf{(r,}t)=\int d\mathbf{v}Ef(\mathbf{x,}t)=\int d\mathbf{v}%
ET_{t,t_{o}}f(\mathbf{x}_{o}\mathbf{,}t_{o}).  \label{4m}
\end{equation}%
\emph{Here} $E\equiv \frac{1}{3}u^{2},$ \emph{with }$\mathbf{u=v-V,}$ \emph{%
while }$P_{o}=P_{o}(t)$\emph{\ is a smooth strictly positive function}$%
\mathbf{.}$\emph{\ Then, the following results follow:}

\emph{1) provided the initial distribution function} $f(\mathbf{x}_{o}%
\mathbf{,}t_{o})$\emph{\ results strictly positive and suitably smooth in
the sense of (\ref{smoothness of f}), it follows that the velocity moments} $%
F_{G}(\mathbf{r},t)=\int\limits_{U}d^{3}vG(\mathbf{x,}t)f(\mathbf{x},t),$
\emph{where} \ $G(\mathbf{x,}t)=1,\mathbf{v,}E\equiv \frac{1}{3}u^{2},$ $%
\mathbf{uu,u}E$ \emph{and }$\mathbf{u=v-V},$ \emph{exist, are continuous }$%
\overline{\Omega }\times I$\emph{\ and suitably smooth in }$\Omega \times I$
\emph{in the sense of the settings} \emph{(\ref{INSE-6c}),(\ref{INSE-7}),(%
\ref{INSE-8}) and (\ref{NFS})}$.$\emph{\ In addition, the kinetic entropy}%
\begin{equation}
S(t)=-\int\limits_{\Gamma }d\mathbf{x}f(\mathbf{x,}t)\ln f(\mathbf{x,}t)
\label{eq.n-2}
\end{equation}%
\emph{exists and is suitably smooth in} $I$ ;

\emph{2) provided the kinetic pressure} $p(\mathbf{r},t)=\int%
\limits_{U}d^{3}vEf(\mathbf{x},t)$ \emph{is suitably prescribed, there
results identically in} $I$
\begin{equation}
S(t)=S(t_{o})  \label{eq.n-3}
\end{equation}%
\emph{(law of entropy conservation).}

For greater clarity, in Sec.2 the relevant aspects of the inverse kinetic
theory previously developed are recalled \cite{Ellero2005}. This is useful
to introduce the Navier-Stoker evolution operator $T_{t,t_{o}}$ and define
the related probability density in the phase space $\Gamma .$ As a
consequence, it is immediate to prove that $T_{t,t_{o}}$ conserves
probability in the same space. Subsequently, in Sec. 3, the kinetic entropy
the $S(t)$ and its time derivative $\partial S(t)/\partial t$ are evaluated.
It is found that by suitably defining the non-negative kinetic pressure ($%
p_{1}$) the entropy production rate can always be set equal to zero in the
whole time interval $I$ in which by assumption a strong solution of INSE
exists. Implications of the result are pointed out.

\section{The Navier-Stokes evolution operator}

In this Section we briefly recall the formulation of the inverse kinetic
theory developed in \cite{Ellero2005}. This is useful in order to identify\
the Navier-Stokes evolution operator $T_{t,t_{o}}$ which determines the
evolution of the relevant kinetic distribution function $f(\mathbf{x},t)$ in
the extended phase space $\Gamma \times I$ ( where $\Gamma \equiv \Omega
\times U$ , with $U\equiv
\mathbb{R}
^{3}$ denoting a suitable "velocity" space).

The result is obtained by requiring the $f(\mathbf{x},t)$ obeys a Vlasov
kinetic equation of the form
\begin{equation}
L(\mathbf{F})f=0,  \label{inverse kinetic equation}
\end{equation}%
where $\mathbf{x}\equiv (\mathbf{r,v})$ is a state vector spanning the phase
space $\Gamma ,$ $L$ the streaming operator
\begin{equation}
L(\mathbf{F})=\frac{\partial }{\partial t}+\frac{\partial }{\partial \mathbf{%
x}}\cdot \left\{ \mathbf{X}\right\}  \label{streaming operator}
\end{equation}%
and $\mathbf{X}$ a vector field of the form $\mathbf{X}(\mathbf{\mathbf{x},}%
t)\mathbf{=}\left\{ \mathbf{v,F}(\mathbf{x},t)\right\} .$ This equation can
formally be written in integral form by introducing the initial kinetic
distribution function $f(\mathbf{x}_{o},t_{o})\equiv f_{o}(\mathbf{x}_{o}),$
defined at the initial time $t_{o}\in I,$ and the flow, i.e., the the
diffeomorphism $\mathbf{x}_{o}\rightarrow \mathbf{x(}t\mathbf{)=}T_{t,t_{o}}%
\mathbf{x}_{o}$ generated by the vector field $\mathbf{X}$, via the
initial-value problem%
\begin{equation}
\left\{
\begin{array}{c}
\frac{d}{dt}\mathbf{x=X}(\mathbf{\mathbf{x},}t) \\
\mathbf{\mathbf{x}}(t_{o})=\mathbf{x}_{o}%
\end{array}%
\right.
\end{equation}%
and its related evolution operator $T_{t,t_{o}}$ (\emph{Navier-Stokes
evolution operator}). This implies that provided the solution of the
initial-value problem exists, is unique and suitably smooth, the Jacobian of
the flow $\mathbf{x}_{o}\rightarrow \mathbf{x(}t\mathbf{),}$ $J\mathbf{(x(}t%
\mathbf{),}t)\equiv \left\vert \frac{\partial \mathbf{x(}t\mathbf{)}}{%
\partial \mathbf{x}_{o}}\right\vert $ is non singular and reads
\begin{equation}
J\mathbf{(x(}t\mathbf{),}t)=\exp \left\{ \int_{t_{o}}^{t}dt^{\prime }\frac{%
\partial }{\partial \mathbf{v(}t^{\prime })}\cdot \mathbf{F(x(}t^{\prime })%
\mathbf{,}t^{\prime }\mathbf{)}\right\} .  \label{Jacobian-1}
\end{equation}%
Thus, the evolution operator, acting on the kinetic distribution function $%
f_{o}\mathbf{(x}_{o})$ results

\begin{equation}
f\mathbf{(x(}t\mathbf{),}t)=T_{t,t_{o}}f_{o}\mathbf{(x}_{o})\equiv f_{o}%
\mathbf{(x}_{o})\exp \left\{ -\int_{t_{o}}^{t}dt^{\prime }\frac{\partial }{%
\partial \mathbf{v(}t^{\prime })}\cdot \mathbf{F(x(}t^{\prime })\mathbf{,}%
t^{\prime }\mathbf{)}\right\} .  \label{integral kinetic equation -0}
\end{equation}

Previously it has been shown \cite{Ellero2006} that the functional form of
the vector field $\mathbf{X}$ and of the "mean-field force" $\mathbf{F}(%
\mathbf{x},t)$ yielding and inverse kinetic theory for INSE can be uniquely
determined under suitable prescriptions. These include, in particular, the
requirements that:

1) the local Maxwellian kinetic distribution function%
\begin{equation}
f_{M}(\mathbf{x,}t;\mathbf{V,}p_{1})=\frac{\rho _{o}}{\left( \pi \right) ^{%
\frac{3}{2}}v_{th}^{3}}\exp \left\{ -X^{2}\right\} ,
\label{local-Maxwellian}
\end{equation}%
[where $X^{2}=\frac{u^{2}}{vth^{2}},$ $v_{th}^{2}=2p_{1}/\rho _{o}$ and $%
\mathbf{u}$ is the relative velocity $\mathbf{u}\mathbb{\equiv }\mathbf{v}-%
\mathbf{V}(\mathbf{r,}t)$] results a particular solution of the inverse
kinetic equation if and only if $\left\{ \rho ,\mathbf{V,}p\right\} $
satisfy INSE;

2) suitable bounce-back boundary conditions are imposed for the kinetic
distribution function on the boundary $\delta \Omega $ \cite{Ellero2005};

3) the moment equations corresponding to the velocity moments $%
M_{G}(r,t)=\int d^{3}vG(\mathbf{x},t)f(\mathbf{x,}t)$ for $G(\mathbf{x},t)=1,%
\mathbf{v/}\rho _{o},E=\frac{1}{3}u^{2}$ coincide with the differential
equations of INSE;

4) the fluid fields $\rho _{o}$ and $\mathbf{V}(\mathbf{r},t)$ are
identified respectively with the velocity moments for $G(\mathbf{x},t)=1,%
\mathbf{v/}\rho _{o};$ similarly, the fluid pressure $p\mathbf{(r,}t)$ is
defined in terms of the kinetic pressure $p_{1}(\mathbf{r,}t)$ [see Eq.(\ref%
{3m})] by requiring%
\begin{equation}
p\mathbf{(r,}t)=p_{1}\mathbf{(r,}t)-P_{o}.
\end{equation}%
It is obvious, in order that $\nabla p=\nabla p_{1},$ that $P_{o}$ can be in
principle an arbitrary strictly positive function independent of $\mathbf{r}$%
. Thus it can always be assumed to be $\forall t\in I$ a smooth function of $%
t$. The resulting form of $\mathbf{F(r,v,}t)$ is recalled in Appendix B. It
implies%
\begin{equation}
\frac{\partial }{\partial \mathbf{v}}\cdot \mathbf{F(x,}t\mathbf{)=}\frac{3}{%
2p_{1}}\left\{ \frac{D}{Dt}p_{1}\mathbf{+\nabla \cdot Q}+\right.
\label{Jacobian-a}
\end{equation}%
\begin{equation*}
\left. +\frac{1}{2p_{1}}\left[ \mathbf{\nabla \cdot }\underline{\underline{%
\mathbf{\Pi }}}\right] \mathbf{\cdot Q}\right\} +\frac{1}{p_{1}}\mathbf{%
u\cdot \nabla \cdot }\underline{\underline{\mathbf{\Pi }}}
\end{equation*}%
where $\frac{D}{Dt}\equiv \frac{\partial }{\partial t}+\mathbf{V\cdot }\frac{%
\partial }{\partial \mathbf{r}}$ is the Lagrangian (or convective)
derivative. For $f\equiv f_{M}$ it becomes%
\begin{equation}
\frac{\partial }{\partial \mathbf{v}}\cdot \mathbf{F(x,}t\mathbf{)=}\frac{3}{%
2p_{1}}\frac{D}{Dt}p_{1}+\frac{1}{p_{1}}\mathbf{u\cdot }\nabla p,
\label{Jacobian-b}
\end{equation}%
which implies that if $f\equiv f_{M}$ $\ $at a given time These expressions,
in particular (\ref{Jacobian-a}) and (\ref{Jacobian-b}), permit to determine
uniquely the Jacobian $J\mathbf{(x(}t\mathbf{),}t)$ and the evolution
operator $T_{t,t_{o}}$. Thus, introducing the normalized kinetic
distribution function
\begin{equation}
\widehat{f}(\mathbf{x},t)=\frac{1}{\rho _{o}}f(\mathbf{x},t)
\end{equation}%
and requiring that the initial kinetic distribution function $\widehat{f}(%
\mathbf{x}_{o},t_{o})\equiv \widehat{f}_{o}(\mathbf{x}_{o})$ results at
least of class $C^{(1)}(\Gamma \times I)$ and summable in $\Gamma $ it
follows
\begin{equation}
d\mathbf{x}(t)\widehat{f}(\mathbf{x}(t),t)=d\mathbf{x}(t_{o})\widehat{f}(%
\mathbf{x}(t_{o}),t_{o})\equiv d\mathbf{x}_{o}\widehat{f}_{o}(\mathbf{x}%
_{o}),  \label{probability conservation}
\end{equation}%
and in particular%
\begin{equation}
\int\limits_{\Gamma }d\mathbf{x}(t)\widehat{f}(\mathbf{x}(t),t)=\int%
\limits_{\Gamma }d\mathbf{x}_{o}\widehat{f}_{o}(\mathbf{x}_{o})=1.
\end{equation}

In order to prove that $\widehat{f}(\mathbf{x}(t),t)$ can be interpreted as
probability density in the next section we intend to establish an H-theorem

\bigskip

\section{Shannon kinetic entropy}

In terms of the Navier-Stokes evolution operator $T_{t,t_{o}}$ and Eq.(\ref%
{Jacobian-a}) [or (\ref{Jacobian-b}) in the case in which $f(\mathbf{x,}t)$
coincides with a local Maxwellian distribution (\ref{local-Maxwellian})] it
is now immediate to evaluate the Shannon kinetic entropy associated to the
kinetic distribution function $f(\mathbf{x,}t),$ namely $S(t)=-\int_{\Gamma
}d\mathbf{x}f(\mathbf{x,}t)\ln f(\mathbf{x,}t).$ Let us assume, for this
purpose that the initial kinetic distribution function $f(\mathbf{x}%
_{o},t_{o})\equiv f_{o}(\mathbf{x}_{o})$ be defined in such a way that it
results strictly positive in $\Gamma $, at least of class $C^{(1)}(\Gamma
\times I)$ and summable in $\Gamma $ so that the Shannon kinetic entropy $%
S(t_{o})=-\int_{\Gamma }d\mathbf{x}_{o}f(\mathbf{x}_{o},t_{o})\ln f(\mathbf{x%
}_{o},t_{o})$ results defined and at least of class $C^{(1)}(I)$. Thanks to
the integral kinetic equation (\ref{integral kinetic equation -0}) and the
condition of conservation (\ref{probability conservation}) it follows that $%
S(t)$ and $S(t_{o})$ are elated by means of the equation:%
\begin{equation}
S(t)=S(t_{o})+\int_{\Gamma }d\mathbf{x}_{o}f_{o}(\mathbf{x}%
_{o})\int_{t_{o}}^{t}dt^{\prime }\frac{\partial }{\partial \mathbf{v(}%
t^{\prime })}\cdot \mathbf{F(x(}t^{\prime })\mathbf{,}t^{\prime }\mathbf{).}
\end{equation}%
Therefore, the entropy production rate $\frac{\partial }{\partial t}S(t)$
results%
\begin{equation}
\frac{\partial }{\partial t}S(t)=\int_{\Gamma }d\mathbf{x}f(\mathbf{x},t)%
\frac{\partial }{\partial \mathbf{v}}\cdot \mathbf{F(x,}t\mathbf{),}
\end{equation}%
where $\frac{\partial }{\partial \mathbf{v}}\cdot \mathbf{F(x,}t\mathbf{)}$
is given or an arbitrary kinetic distribution function by Eq.(\ref%
{Jacobian-a}) [or (\ref{Jacobian-b}) for the Maxwellian case]. It follows
\begin{eqnarray}
&&\text{ \ \ \ \ \ \ \ \ \ \ \ \ \ }\left. \frac{\partial }{\partial t}S(t)=%
\frac{3}{2}\int_{\Gamma }d\mathbf{x}\frac{1}{P_{o}+p(\mathbf{r},t)}f(\mathbf{%
x},t)\right.  \\
&&\text{ \ \ \ \ \ \ \ \ \ }\left. \left\{ \frac{\partial }{\partial t}%
P_{o}(t)+\frac{D}{Dt}p\mathbf{+\nabla \cdot Q+}\frac{1}{2p_{1}}\left[
\mathbf{\nabla \cdot }\underline{\underline{\mathbf{\Pi }}}\right] \mathbf{%
\cdot Q}\right\} .\right.
\end{eqnarray}%
Since $P_{o}+p(\mathbf{r},t)$ and $f(\mathbf{x},t)$ are strictly positive,
we can always define $P_{o}(t)$ so that in the finite time interval $I$
there results identically

\begin{equation}
\frac{\partial }{\partial t}P_{o}(t)=\frac{-\int_{\Omega }d\mathbf{r}\frac{1%
}{P_{o}(t)+p(\mathbf{r},t)}\left[ \frac{D}{Dt}p\mathbf{+\nabla \cdot Q+}%
\frac{1}{2}\left[ \mathbf{\nabla \cdot }\underline{\underline{\mathbf{\Pi }}}%
\right] \mathbf{\cdot Q}\right] }{\int_{\Omega }d\mathbf{r}\frac{1}{%
P_{o}(t)+p(\mathbf{r},t)}}.  \label{vanishing of entropy production}
\end{equation}%
In the case in which the initial condition $f_{o}(\mathbf{x}_{o})$ coincides
identically with $f_{M}$ in the whole phase space $\Gamma $ it follows in
particular%
\begin{equation}
\frac{\partial }{\partial t}P_{o}(t)=\frac{-\int_{\Omega }d\mathbf{r}\frac{1%
}{P_{o}(t)+p(\mathbf{r},t)}\frac{D}{Dt}p}{\int_{\Omega }d\mathbf{r}\frac{1}{%
P_{o}(t)+p(\mathbf{r},t)}}.  \label{vanishing-entropy prod -f_M}
\end{equation}%
Therefore, condition (\ref{vanishing of entropy production}) [or (\ref%
{vanishing-entropy prod -f_M}) in the Maxwellian case] implies that in the
same time interval $I$ the entropy production rate must vanish identically,
i.e.,
\begin{equation}
\frac{\partial }{\partial t}S(t)\equiv 0.  \label{H-theorem}
\end{equation}%
We stress that Eq.(\ref{H-theorem}) holds, in principle, for an arbitrary
initial condition $P_{o}(t_{o})=P_{oo}>0$ with $P_{oo}$ suitably large$.$
Therefore, the kinetic pressure $p_{1},$ given by Eqs.(\ref{3m}) and (\ref%
{4m}), remains still non-unique since it is determined in terms of Eq.(\ref%
{vanishing of entropy production}) only up to an arbitrary positive constant
$P_{oo}.$

It follows that the Shannon entropy for the kinetic distribution function $f(%
\mathbf{x,}t)$ results always conserved by imposing a suitable prescription
on the the time-dependent part of the kinetic pressure $P_{o}(t)$. Since the
latter is unrelated to the physical observables (i.e., the fluid fields) the
constraint condition imposed on the kinetic pressure [respectively (\ref%
{vanishing of entropy production}) or (\ref{vanishing-entropy prod -f_M})]
can always be satisfied. As a consequence, with such prescriptions the
normalized kinetic distribution function $\widehat{f}(\mathbf{x},t)$ can be
interpreted as probability density.

\subsection{Conclusions}

In this paper the condition of positivity of the kinetic distribution
function $f(\mathbf{x},t)$ which characterizes the inverse kinetic theory
recently developed for the incompressible Navier-Stokes equations has been
investigated \cite{Ellero2005,Ellero2006}. We have proven that the Shannon
entropy is exactly conserved for arbitrary kinetic distribution function,
provided the kinetic pressure is suitably defined and the initial kinetic
distribution function results positive definite and suitably regular. As
indicated, these conditions can always be satisfied without imposing any
constraint on the physical observables, here represented by the fluid fields
\ $\left\{ \rho _{o},\mathbf{V,}p\right\} .$

The conclusion applies in principle to arbitrary, suitable smooth in the
sense (\ref{NFS}), strong solutions of INSE which are defined in three
dimensional internal domains of $%
\mathbb{R}
^{3}.$ Assuming, mass conservation and no-slip boundary conditions (i.e.,
Dirichlet boundary conditions) on the boundary $\delta \Omega ,$ the same
result holds also for non-isolated systems characterized by moving
boundaries. In addition, arbitrary volume forces which satisfy (\ref{INSE-7}%
),(\ref{INSE-8}) or analogous surface forces obtained by applying a
non-uniform pressure on the boundary $\delta \Omega $, can be included.

An immediate consequence of the H-theorem here obtained is the possibility
of imposing the maximum entropy principle in order to determine the initial
kinetic distribution function $f_{o}(\mathbf{x})$, i.e., requiring the
variational equation $\delta S(f_{o})=0$ subject to suitable constraint
equations. Thus, for example, the local Maxwellian distribution (\ref%
{local-Maxwellian}) is obtained by imposing solely the constraints provided
by the moments (\ref{1m}),(\ref{2m}),(\ref{3m}) and (\ref{4m}), to be
considered as prescribed. However, in principle, the variational principle
can also be used to determine non-Maxwellian initial distributions \cite%
{Jaynes1957}.

\bigskip These results appear significant both from the mathematical
viewpoint and the physical interpretation of the theory.

\textbf{ACKNOWLEDGEMENTS} The research was developed in the framework of the
PRIN Research Project "Modelli della teoria cinetica matematica nello studio
dei sistemi complessi nelle scienze applicate" (Italian Ministry of
University and Research)

\bigskip

\section{Appendix A: INSE}

The incompressible Navier-Stokes equations (INSE) are defined by the
following set of PDE's and inequalities for the fluid fields $\left\{ \rho ,%
\mathbf{V,}p\right\} $

\begin{eqnarray}
\frac{\partial }{\partial t}\rho +\nabla \cdot \left( \rho \mathbf{V}\right)
&=&0,  \label{INSE-1} \\
\text{ \ \ \ \ \ \ \ \ \ \ \ \ \ \ \ \ \ \ \ \ \ \ \ \ }\rho \frac{D}{Dt}%
\mathbf{V}+\mathbf{\nabla }p+\mathbf{f}-\mu \nabla ^{2}\mathbf{V} &=&\mathbf{%
0},  \label{INSE-2} \\
\nabla \cdot \mathbf{V} &=&0,  \label{INSE-3} \\
\rho (\mathbf{r,}t) &>&0,  \label{INSE-4} \\
p(\mathbf{r,}t) &\geq &0,  \label{INSE-5} \\
\rho (\mathbf{r,}t\mathbb{)} &=&\rho _{o}>0.  \label{INSE-6b}
\end{eqnarray}%
The first three equations (\ref{INSE-1}),(\ref{INSE-2}) and (\ref{INSE-3}),
denoting respectively the continuity, forced Navier-Stokes and isochoricity
equations, are assumed to be satisfied in the open three-dimensional set $%
\Omega \subseteq
\mathbb{R}
^{3}$ (fluid domain) and in a possibly bounded time interval $I\subset
\mathbb{R}
,$ while the last three inequalities, (\ref{INSE-4})-(\ref{INSE-6b}) apply
also in the closure of the fluid domain $\overline{\Omega }\equiv \Omega
\cup \delta \Omega .$ Here the notation is standard\cite{Ellero2005}. Hence $%
\frac{D}{Dt}=\frac{\partial }{\partial t}+\mathbf{V\cdot \nabla }$ and $\mu
\equiv \nu \rho _{o}>0$ is the constant fluid viscosity, with $\nu $ the
related kinematic viscosity. The volume force density $\mathbf{f(r,}t)$
acting on the fluid element by assumption is taken in the functional setting
(\ref{INSE-7}),(\ref{INSE-8}) and (\ref{NFS}).\ Consequently, the fluid
fields $\left\{ \mathbf{V}(\mathbf{\mathbf{r,}}t),p(\mathbf{r,}t)\right\} $
are required to satisfy the regularity conditions (\ref{NFS}). The
initial-boundary value problem for INSE is defined as follows. The initial
condition is defined by imposing
\begin{eqnarray}
\text{ \ \ \ \ \ \ \ \ \ \ \ \ \ \ \ \ \ \ \ \ \ \ \ \ \ \ \ \ \ \ \ \ \ \ }%
\rho (\mathbf{r,}t_{o}\mathbb{)} &=&\rho _{o}>0 \\
p(\mathbf{r,}t_{o}) &=&p_{o}(\mathbf{r}), \\
\mathbf{V}(\mathbf{\mathbf{r,}}t_{o}) &=&\mathbf{V}_{o}(\mathbf{\mathbf{r}}),
\end{eqnarray}%
where $\left\{ \mathbf{V}_{o}(\mathbf{\mathbf{r}}),p_{o}(\mathbf{r})\right\}
$ \ belong to the functional class%
\begin{equation}
\left\{
\begin{array}{l}
\mathbf{V}_{o}(\mathbf{\mathbf{r}}),p_{o}(\mathbf{r})\in C^{(0)}(\overline{%
\Omega }), \\
\mathbf{V}_{o}(\mathbf{\mathbf{r}}),p_{o}(\mathbf{r})\in C^{(2)}(\Omega ),%
\end{array}%
\right.
\end{equation}%
and moreover satisfy respectively the isochoricity condition (\ref{INSE-3})
and the Poisson equation
\begin{equation}
\nabla ^{2}p_{o}(\mathbf{r})=-\nabla \cdot \left\{ \rho _{o}\mathbf{V}_{o}%
\mathbf{\cdot \nabla V}_{o}+\mathbf{f}(\mathbf{r,}t_{o})\right\} .
\end{equation}%
The boundary conditions can be specified, for example, by means of the
Dirichlet boundary conditions (which for the velocity are usually denoted as
no-slip boundary conditions), i.e., letting $\forall t\in I$ and imposing in
each point $\mathbf{\mathbf{r}}_{W}$ of the the boundary $\delta \Omega $
\begin{eqnarray}
\text{ \ \ \ \ \ \ \ \ \ \ \ \ \ \ \ \ \ \ \ \ \ \ \ \ \ \ \ \ \ \ \ \ \ }%
\rho (\cdot \mathbf{,}t\mathbb{)} &=&\rho _{o}>0, \\
p(\cdot \mathbf{,}t) &=&p_{W}(\cdot \mathbf{,}t), \\
\mathbf{V}(\cdot \mathbf{\mathbf{,}}t) &=&\mathbf{V}_{W}(\cdot \mathbf{%
\mathbf{,}}t).
\end{eqnarray}%
Here $\left\{ \mathbf{V}_{W}(\mathbf{\mathbf{r,}}t),p_{W}(\mathbf{r,}%
t)\right\} $ denote respectively the velocity and the pressure at an
arbitrary point $\mathbf{r}_{W}$ belonging to the boundary $\delta \Omega .$
fields, both required to belong to the same functional class (\ref{NFS}).

\section{Appendix B: mean-field force}

For a generic (i.e., non-Maxwellian) distribution function $f(\mathbf{x,}t),$
the mean-field force $\mathbf{F}$ reads $\mathbf{F(x,}t)=\mathbf{F}_{0}%
\mathbf{(x,}t)+\mathbf{F}_{1}\mathbf{(x,}t),$ where $\mathbf{F}_{0}$ and $%
\mathbf{F}_{1}$ are the vector fields:

\begin{equation}
\mathbf{F}_{0}\mathbf{(x,}t)=\frac{1}{\rho _{o}}\left[ \mathbf{\nabla \cdot }%
\underline{\underline{\mathbf{\Pi }}}-\mathbf{\nabla }p_{1}-\mathbf{f}\right]
+\mathbf{a+}\nu \nabla ^{2}\mathbf{V,}  \label{F0 non-maxwellian case}
\end{equation}%
\begin{equation}
\mathbf{F}_{1}\mathbf{(x,}t)=\frac{1}{2p_{1}}\mathbf{u}\left\{ \frac{D}{Dt}%
p_{1}\mathbf{+\nabla \cdot Q}+\right.  \label{F1 non-Maxwellian case}
\end{equation}%
\begin{equation*}
\left. +\frac{1}{2p_{1}}\left[ \mathbf{\nabla \cdot }\underline{\underline{%
\mathbf{\Pi }}}\right] \mathbf{\cdot Q}\right\} +\frac{v_{th}^{2}}{2p_{1}}%
\mathbf{\nabla \cdot }\underline{\underline{\mathbf{\Pi }}}\left\{ X^{2}-%
\frac{3}{2}\right\} .
\end{equation*}%
where $X^{2}=\frac{\mathbf{u}^{2}}{vth^{2}}$ and $v_{th}^{2}=2p_{1}/\rho
_{o}.$ Here $\mathbf{Q}$ and \underline{\underline{$\mathbf{\Pi }$}} are the
velocity-moments $\mathbf{Q}=\int d^{3}v\mathbf{u}\frac{u^{2}}{3}f,$ $%
\underline{\underline{\mathbf{\Pi }}}=\int d^{3}v\mathbf{uu}f,$ while $%
\mathbf{f}$ denotes the volume force density acting on the fluid element and
finally $\nu >0$ is the constant kinematic viscosity. In particular, for the
Maxwellian kinetic equilibrium (\ref{local-Maxwellian}) there results $%
\underline{\underline{\mathbf{\Pi }}}=p_{1}\underline{\underline{\mathbf{1}}}%
,\mathbf{Q=0}$. Moreover, $\mathbf{a}$ is the convective term which
according to Ref.\cite{Ellero2006} is uniquely defined and reads $\mathbf{a=}%
\frac{1}{2}\mathbf{u\cdot \nabla V}+\frac{1}{2}\nabla \mathbf{V\cdot u.}$

\end{document}